\documentclass[10pt,twocolumn]{article}
\usepackage{graphicx}
\usepackage{amsmath,amssymb,times,cite}

\newcommand{\bm}[1]{\mathbf{#1}} 

\newcommand\T{{\mathpalette\raiseT\intercal}}
\newcommand\raiseT[2]{%
\setbox0\hbox{$#1{#2}$}\raise\dp0\box0}

\usepackage{algorithm,algorithmic}
\makeatletter
\newcommand{\ALOOP}[1]{\ALC@it\algorithmicloop\ #1%
  \begin{ALC@loop}}
\newcommand{\ENDALOOP}{\end{ALC@loop}\ALC@it\algorithmicendloop}

\makeatother

\usepackage{booktabs}
\usepackage[table]{xcolor}

\definecolor{lightgray}{gray}{0.9}
\definecolor{lightblue}{rgb}{0.93,0.95,1.0}

\title{\Large\textbf{Melanoma Detection using Adversarial Training and Deep Transfer Learning}}
\author{Hasib Zunair and A. Ben Hamza \\
Concordia University, Montreal, QC, Canada
}
\date{}

\begin{document}
\maketitle

\begin{abstract}
Skin lesion datasets consist predominantly of normal samples with only a small percentage of abnormal ones, giving rise to the class imbalance problem. Also, skin lesion images are largely similar in overall appearance owing to the low inter-class variability. In this paper, we propose a two-stage framework for automatic classification of skin lesion images using adversarial training and transfer learning toward melanoma detection. In the first stage, we leverage the inter-class variation of the data distribution for the task of conditional image synthesis by learning the inter-class mapping and synthesizing under-represented class samples from the over-represented ones using unpaired image-to-image translation. In the second stage, we train a deep convolutional neural network for skin lesion classification using the original training set combined with the newly synthesized under-represented class samples. The training of this classifier is carried out by minimizing the focal loss function, which assists the model in learning from hard examples, while down-weighting the easy ones. Experiments conducted on a dermatology image benchmark demonstrate the superiority of our proposed approach over several standard baseline methods, achieving significant performance improvements. Interestingly, we show through feature visualization and analysis that our method leads to context based lesion assessment that can reach an expert dermatologist level.
\end{abstract}


\section{Introduction}
Melanoma is one of the most aggressive forms of skin cancer~\cite{Saito:2018,Siegel:2019}. It is diagnosed in more than 132,000 people worldwide each year, according to the World Health Organization. Hence, it is essential to detect melanoma early before it spreads to other organs in the body and becomes more difficult to treat.

While visual inspection of suspicious skin lesions by a dermatologist is normally the first step in melanoma diagnosis, it is generally followed by dermoscopy imaging for further analysis. Dermoscopy is a noninvasive imaging procedure that acquires a magnified image of a region of the skin at a very high resolution to clearly identify the spots on the skin~\cite{binder1995epiluminescence}, and helps identify deeper levels of skin, providing more details of the lesions. Moreover, dermoscopy provides detailed visual context of regions of the skin and has proven to enhance the diagnostic accuracy of a naked eye examination, but it is costly, error prone, and achieves only average sensitivity in detecting melanoma~\cite{ganster2001automated}. This has triggered the need for developing more precise computer-aided diagnostics systems that would assist in early detection of melanoma from dermoscopy images. Despite significant strides in skin lesion recognition, melanoma detection remains a challenging task due to various reasons, including the high degree of visual similarity (i.e. low inter-class variation) between malignant and benign lesions, making it difficult to distinguish between melanoma and non-melanoma skin lesions during the diagnosis of patients. Also, the contrast variability and boundaries between skin regions owing to image acquisition make automated detection of melanoma an intricate task. In addition to the high intra-class variation of melanoma's color, texture, shape, size and location in dermoscopic images~\cite{cheng2008skin}, there are also artifacts such as hair, veins, ruler marks, illumination variation, and color calibration charts that usually cause occlusions and blurriness, further complicating the situation~\cite{ZLiu:15}.

Classification of skin lesion images is a central topic in medical imaging, having a relatively extensive literature. Some of the early methods for classifying melanoma and non-melanoma skin lesions have focused mostly on low-level computer vision approaches, which involve hand-engineering features based on expert
knowledge such as color \cite{cheng2008skin}, shape \cite{mishra2016overview} and texture \cite{ballerini2013color, tommasi2006melanoma}. By leveraging feature selection, approaches that use mid-level computer vision techniques have also been shown to achieve improved detection performance~\cite{celebi2007methodological}. In addition to ensemble classification based techniques\cite{schaefer2014ensemble}, other methods include two-stage approaches, which usually involve segmentation of skin lesions, followed by a classification stage to further improve detection performance~\cite{tommasi2006melanoma,ganster2001automated,celebi2007methodological}. However, hand-crafted features often lead to unsatisfactory results on unseen data due to high intra-class variation and visual similarity, as well as the presence of artifacts in dermoscopy images. Moreover, such features are usually designed for specific tasks and do not generalize across different tasks.

Deep learning has recently emerged as a very powerful way to hierarchically find abstract patterns using large amounts of training data. The tremendous success of deep neural networks in image classification, for instance, is largely attributed to open source software, inexpensive computing hardware, and the availability of large-scale datasets~\cite{krizhevsky2012imagenet}. Deep learning has proved valuable for various medical image analysis tasks such as classification and segmentation~\cite{ronneberger2015u,roth2014new,anthimopoulos2016lung,matsunaga2017image, codella2017deep,gutman2016skin}. In particular, significant performance gains in melanoma recognition have been achieved by leveraging deep convolutional neural networks in a two-stage framework~\cite{yu2016automated}, which uses a fully convolutional residual network for skin lesion segmentation and a very deep residual network for skin lesion classification. However, the issues of low inter-class variation and class imbalance of skin lesion image datasets severely undermine the applicability of deep learning to melanoma detection~\cite{shie2015transfer,yu2016automated}, as they often hinder the model's ability to generalize, leading to over-fitting~\cite{shin2016deep}. In this paper, we employ conditional image synthesis without paired images to tackle the class imbalance problem by generating synthetic images for the minority class. Built on top of generative adversarial networks (GANs)~\cite{goodfellow2014generative}, several image synthesis approaches, both conditional \cite{nie2017medical} and unconditional \cite{frid2018synthetic}, have been recently adopted for numerous medical imaging tasks, including melanoma detection~\cite{yi2019generative,costa2017end,zhang2019skrgan}. Also, approaches that enable the training of diverse models based on distribution matching with both paired and unpaired data were introduced in~\cite{zhu2017unpaired, isola2017image, liu2017unsupervised, lamb2017gibbsnet}. These approaches include image translation from CT-PET~\cite{ben2017virtual}, CS-MRI~\cite{yang2017dagan}, MR-CT~\cite{wolterink2017deep}, XCAT-CT~\cite{russ2019synthesis} and H\&E staining in histopathology~\cite{shaban2019staingan, de2018stain}. In~\cite{bissoto2018skin,ali2019data}, image synthesis models that synthesize images from noise were developed in an effort to improve melanoma detection. However, Cohen \textit{et al.}~\cite{cohen2018distribution} showed that the training schemes used in several domain adaptation methods often lead to a high bias and may result in hallucinating features (e.g. adding or removing tumors leading to a semantic change). This is due in large part to the source or target domains consisting of over- or under-represented samples during training (e.g. source domain composed of 50\% malignant images and 50\% benign; or target domain composed of 20\% malignant and 80\% benign images).

In this paper, we introduce MelaNet, a deep neural network based framework for melanoma detection, to overcome the aforementioned issues. Our approach mitigates the bias problem~\cite{cohen2018distribution}, while improving detection performance and reducing over-fitting. The proposed MelaNet framework consists of two integrated stages. In the first stage, we generate synthetic dermoscopic images for the minority class (i.e. malignant images) using unpaired image-to-image translation in a bid to balance the training set. These additional images are then used to boost training. In the second stage, we train a deep convolutional neural network classifier by minimizing the focal loss function, which assists the classification model in learning from hard examples, while down-weighting the easy ones. The main contributions of this paper can be summarized as follows:
\begin{itemize}
\item We propose an integrated deep learning based framework, which couples adversarial training and transfer learning to jointly address inter-class variation and class imbalance for the task of skin lesion classification.
\item We train a deep convolutional network by iteratively minimizing the focal loss function, which assists the model in learning from hard examples, while down-weighting the easy ones.
\item We show experimentally on a dermatology image analysis benchmark significant improvements over several baseline methods for the important task of melanoma detection.
\item We show how our method enables visual discovery of high activations for the regions surrounding the skin lesion, leading to context based lesion assessment that can reach an expert dermatologist level.
\end{itemize}

The rest of this paper is organized as follows. In Section 2, we introduce a two-stage approach for melanoma detection using conditional image synthesis from benign to malignant in an effort to mitigate the effect caused by class imbalance, followed by training a deep convolutional neural network via iterative minimization of the focal loss function in order to learn from hard examples. We also discuss in detail the major components of our approach, and summarize its main algorithmic steps. In Section 3, experiments performed on a dermatology image analysis datasets are presented to demonstrate the effectiveness of the proposed approach in comparison with baseline methods. Finally, we conclude in Section 4 and point out future work directions.

\section{Method} \label{Method}
In this section, we describe the main components and algorithmic steps of the proposed approach to melanoma detection.

\subsection{Conditional Image Synthesis}
In order to tackle the challenging issue of low inter-class variation in skin lesion datasets~\cite{yu2016automated, shin2016deep}, we partition the inter-classes into two domains for conditional image synthesis with the goal to generate malignant lesions from benign lesions. This data generation process for the malignant minority class is performed in an effort to mitigate the class imbalance problem, as it is relatively easy to learn a transformation with given prior knowledge or conditioning for a narrowly defined task~\cite{de2018stain,wolterink2017deep}. Also, using unconditional image synthesis to generate data of a target distribution from noise often leads to artifacts and may result in training instabilities~\cite{zhao2019compression}. In recent years, various methods based on generative adversarial networks (GANs) have been used to tackle the conditional image synthesis problem, but most of them use paired training data for image-to-image translation~\cite{kazeminia2018gans}, which requires the generation of a new image that is a controlled modification of a given image. Due to the unavailability of datasets consisting of paired examples for melanoma detection, we use cycle-consistent adversarial networks (CycleGAN), a technique that involves the automatic training of image-to-image translation models without paired examples~\cite{zhu2017unpaired}. These models are trained in an unsupervised fashion using a collection of images from the source and target domains. CycleGAN is a framework for training image-to-image translation models by learning mapping functions between two domains using the GAN model architecture in conjunction with cycle consistency. The idea behind cycle consistency is to ward off the learned mappings between these two domains from contradicting each other.

\begin{figure}[!htb]
\centering
\includegraphics[scale=.54]{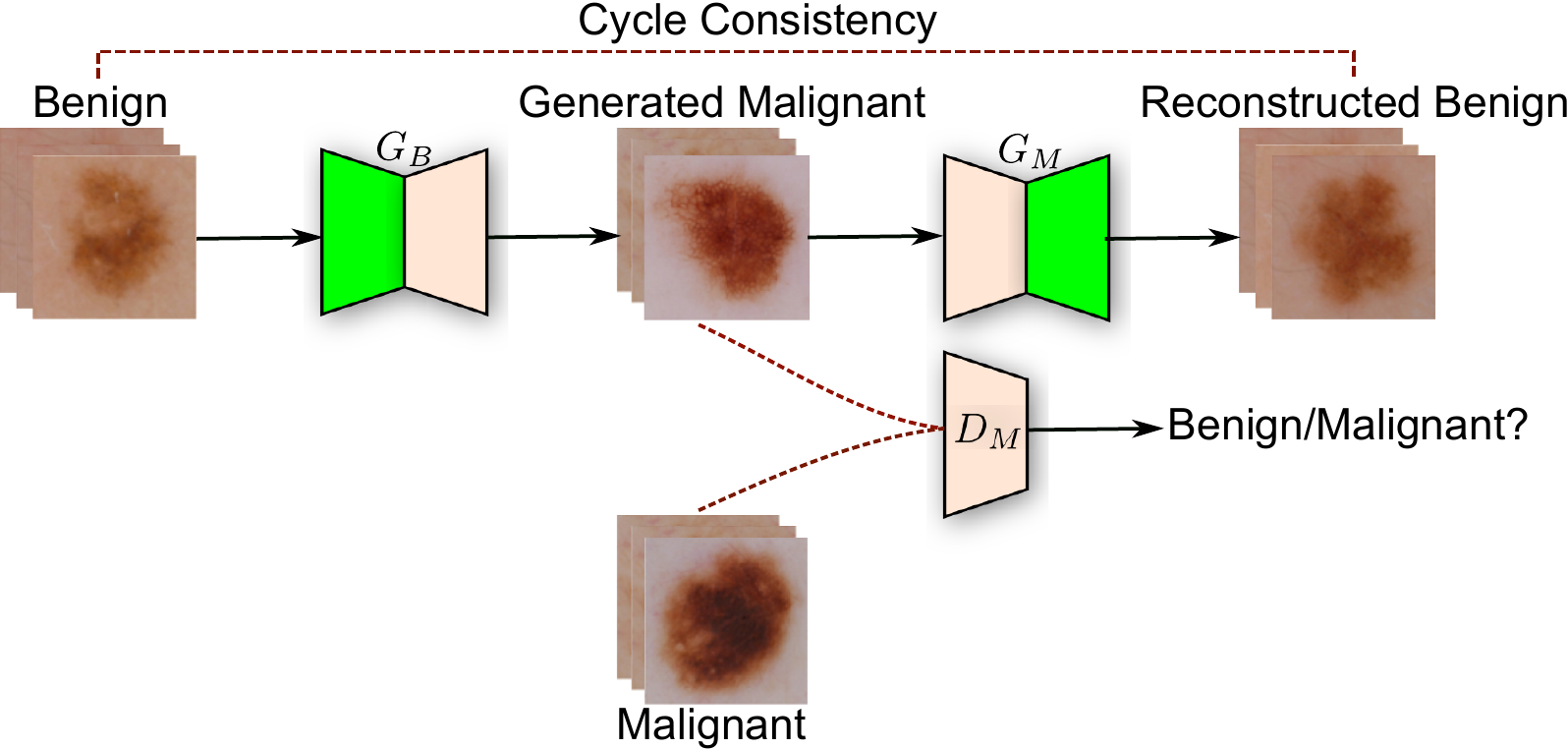}
\caption{Illustration of the generative adversarial training process for unpaired image-to-image translation. Lesions are translated from benign to malignant and then back to benign to ensure cycle consistency in the forward pass. The same procedure is applied in the backward pass from malignant to benign.}
\label{Fig:MelaNet}
\end{figure}

Given two image domains $B$ and $M$ denoting benign and malignant, respectively, the CycleGAN framework aims to learn to translate images of one type to another using two generators $G_{B}: B\rightarrow M$ and $G_{M}: M\rightarrow B$, and two discriminators $D_{M}$ and $D_{B}$, as illustrated in Figure \ref{Fig:MelaNet}. The generator $G_{B}$ (resp. $G_{M}$) translates images from benign to malignant (resp. malignant to benign), while the discriminator $D_{M}$ (resp. $D_{B}$) scores how real an image of $M$ (resp. $B$) looks. In other words, these discriminator models are used to determine how plausible the generated images are and update the generator models accordingly. The objective function of CycleGAN is defined as
\begin{equation}
\begin{split}
\mathcal{L}(G_{B}, G_{M}, D_{M}, D_{B}) &= \mathcal{L}_{GAN}(G_{B}, D_{M}, B, M) \\
&\quad+ \mathcal{L}_{GAN}(G_{M}, D_{B}, M, B) \\
&\quad+ \lambda\,\mathcal{L}_{cyc}(G_{B}, G_{M}),
\end{split}
\label{eq:cyclegan_loss}
\end{equation}
which consists of two adversarial loss functions and a cycle consistency loss function regularized by a hyper-parameter $\lambda$ that controls the relative importance of these loss functions~\cite{zhu2017unpaired}. The first adversarial loss is given by
\begin{equation}
\begin{split}
\mathcal{L}_{GAN}(G_{B},D_{M},B,M) &= \mathbb{E}_{m\sim p_{\text{data}}(m)}[\log D_{M}(m)] \\
&\hspace*{-.4in} + \mathbb{E}_{b\sim p_{\text{data}}(b)}[\log( 1 - D_{M}(G_{B}(b)))],
\end{split}
\label{eq:advloss}
\end{equation}
where the generator $G_B$ tries to generate images $G_{B}(b)$ that look similar to malignant images, while $D_M$ aims to distinguish between generated samples $G_{B}(b)$ and real samples $m$. During the training, as $G_{B}$ generates a malignant lesion, $D_{M}$ verifies if the translated image is actually a real malignant lesion or a generated one. The data distributions of benign and malignant are $p_{\text{data}}(b)$ and $p_{\text{data}}(m)$, respectively. Similarly, the second adversarial loss is given by
\begin{equation}
\begin{split}
\mathcal{L}_{GAN}(G_{M},D_{B},M,B) &= \mathbb{E}_{b\sim p_{\text{data}}(b)}[\log D_{B}(b)] \\
&\hspace*{-.4in} + \mathbb{E}_{m \sim p_{\text{data}}(m)}[\log( 1 - D_{B}(G_{M}(m)))],
\end{split}
\label{eq:advloss}
\end{equation}
where $G_M$ takes a malignant image $m$ from $M$ as input, and tries to generate a realistic image $G_{M}(m)$ in $B$ that tricks the discriminator $D_B$. Hence, the goal of $G_{M}$ is to generate a benign lesion such that it fools the discriminator $D_{B}$ to label it as a real benign lesion.

\noindent The third loss function is the cycle consistency loss given by
\begin{equation}
\begin{split}
\mathcal{L}_{cyc} (G_{B}, G_{M}) &= \mathbb{E}_{b \sim p_{\text{data}} (b)} [\|G_{M}(G_{B}(b)) - b\|_1] \\
&\,\,+ \mathbb{E}_{m \sim p_{\text{data}} (m)} [\|G_{B}(G_{M}(m)) - m\|_1],
\end{split}
\label{eq:cycleloss}
\end{equation}
which basically quantifies the difference between the input image and the generated one using the $\ell_1$-norm. The idea of the cycle consistency loss it to enforce $G_{M}(G_{B}(b))\approx b$ and $G_{B}(G_{M}(m))\approx m$. In other words, the objective of CycleGAN is to learn two bijective generator mappings by solving the following optimization problem
\begin{equation}
G_{B}^{\ast}, G_{M}^{\ast}=\arg\min_{G_{B},G_{M}}\max_{D_{B},D_{M}}\mathcal{L}(G_{B}, G_{M}, D_{M}, D_{B}).
\end{equation}
We adopt the U-Net architecture~\cite{ronneberger2015u} for the generators and PatchGAN~\cite{isola2017image} for the discriminators. The U-Net architecture consists of an encoder subnetwork and decoder subnetwork that are connected by a bridge section, while PatchGAN is basically a convolutional neural network classifier that determines whether an image patch is real or fake.

\subsection{Pre-trained Model Architecture} \label{VGG-GAP}
\begin{figure}[!htb]
\centering
\includegraphics[scale=.38]{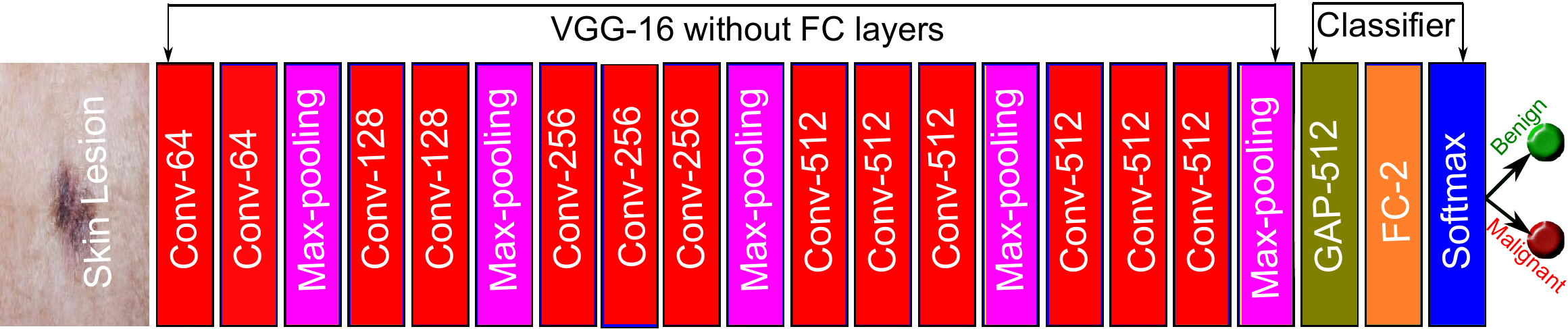}
\caption{VGG-GAP architecture with a GAP layer, followed by an FC layer that in turn is fed into a softmax layer of two units.}
\label{Fig:vgg-gap}
\end{figure}

Due to limited training data, it is standard practice to leverage deep learning models that were pre-trained on large datasets~\cite{yosinski2014transferable}. The proposed melanoma classification model uses the pre-trained VGG-16 convolutional neural network without the fully connected (FC) layers, as illustrated in Figure~\ref{Fig:vgg-gap}. The VGG-16 network consists of 16 layers with learnable weights: 13 convolutional layers, and 3 fully connected layers~\cite{simonyan2014very}. As shown in Figure~\ref{Fig:vgg-gap}, the proposed architecture, dubbed VGG-GAP, consists of five blocks of convolutional layers, followed by a global average pooling (GAP) layer. Each of the first and second convolutional blocks is comprised of two convolutional layers with 64 and 128 filters, respectively. Similarly, each of the third, fourth and fifth convolutional blocks consists of three convolutional layers with 256, 512, and 512 filters, respectively. The GAP layer, which is widely used in classification tasks, computes the average output of each feature map in the previous layer and helps minimize overfitting by reducing the total number of parameters in the model. GAP turns a feature map into a single number by taking the average of the numbers in that feature map. Similar to max pooling layers, GAP layers have no trainable parameters and are used to reduce the spatial dimensions of a three-dimensional tensor. The GAP layer is followed by a single FC layer with a softmax function (i.e. a dense softmax layer of two units for the binary classification case) that yields the probabilities of predicted classes.

Since we are addressing a binary classification problem with imbalanced data, we learn the weights of the VGG-GAP network by minimizing the focal loss function~\cite{lin2017focal} defined as
\begin{equation}
\text{FL}(p_{t}) = - \alpha_{t}(1- p_t)^\gamma\log(p_t),
\label{eq:fl}
\end{equation}
where $p_t$ and $\alpha_t$ are given by
$$
p_{t}=\begin{cases}
p & \text{if $y=1$},\\
1-p & \text{otherwise}
\end{cases}
\quad\text{and}\quad
\alpha_{t}=\begin{cases}
\alpha & \text{if $y=1$},\\
1-\alpha & \text{otherwise},
\end{cases}
$$
with $y\in\{-1,1\}$ denoting the ground truth for negative and positive classes, and $p\in[0,1]$ denoting the model's predicted probability for the class with label $y=1$. The weight parameter $\alpha\in[0,1]$ balances the importance of positive and negative labeled samples, while the nonnegative tunable focusing parameter $\gamma$ smoothly adjusts the rate at which easy examples are down-weighted. Note that when $\gamma=0$, the focal loss function reduces to the cross-entropy loss. A positive value of the focusing parameter decreases the relative loss for well-classified examples, focusing more on hard, misclassified examples.

Intuitively, the focal loss function penalizes hard-to-classify examples. It basically down-weights the loss for well-classified examples so that their contribution to the total loss is small even if their number is large.

\subsection{Data Preprocessing and Augmentation} \label{Data Preprocessing and Augmentation}
In order to achieve faster convergence, feature standardization is usually performed, i.e. we rescale the images to have values between 0 and 1. Given a data matrix $\bm{X}=(\bm{x}_{1},\ldots,\bm{x}_{n})^{\T}$, the standardized feature vector is given by
\begin{equation}
\bm{z}_{i} = \frac{\bm{x}_{i} - \min(\bm{x}_{i})}{\max(\bm{x}_{i}) - \min(\bm{x}_{i})},\quad i=1,\ldots,n,
\label{eq:std}
\end{equation}
where $\bm{x}_{i}$ is the $i$-th input data point, denoting a row vector. It is important to note that in our approach, no domain specific or application specific pre-processing or post-processing is employed.

On the other hand, data augmentation is usually carried out on medical datasets to improve performance in classification tasks~\cite{frid2018synthetic, araujo2017classification}. This is often done by creating modified versions of the input images in a dataset through random transformations, including horizontal
and vertical flip, Gaussian noise, brightness and zoom augmentation, horizontal and vertical shift, sampling noise once per pixel, color space conversion, and rotation.

We do not perform on-the-fly data augmentation (random) during training, as it may add an unnecessary layer of complexity to training and evaluation. When designing our configurations, we first augment the data offline and then we train the classifier using the augmented data. Also, we do not apply data augmentation in the proposed two-stage approach, as it would not give us an insight on which of the two approaches has more contribution in the performance (data augmentation or image synthesis?). Hence, we keep these two configurations independent from each other.

\subsection{Algorithm}
The main algorithmic steps of our approach are summarized in Algorithm~\ref{algo:algoooo}. The input is a training set consisting of skin lesion dermoscopic images, along with their associated class labels. In the first stage, the different classes are grouped together (e.g. for binary classification, we have two groups), and we resize each image to $256\times 256\times 3$. Then, we balance the inter-class data samples by performing undersampling. We train CycleGAN to learn a function of the interclass variation between the two groups, i.e. we learn a transformation between melanoma and non-melanoma lesions. We apply CycleGAN to the over-represented class samples in order to synthesize the target class samples (i.e. under-represented class). After this transformation is applied, we acquire a balanced dataset, composed of original training data and generated data. In the second stage, we employ the VGG-GAP classifier with the focal loss function. Finally, we evaluate the trained model on the test set to generate the predicted class labels.

\begin{algorithm}
  \caption{MelaNet classifier}
  \label{algo:algoooo}
  \begin{algorithmic}[1]
    \REQUIRE Training set $\mathcal{D}=\{(\bm{I}_1,y_1),\dots,(\bm{I}_n,y_n)\}$ of dermoscopic images, where $y_i$ is a class label of the input $\bm{I}_i$.
    \ENSURE Vector $\bm{\hat{y}}$ containing predicted class labels.
    \FOR{$i=1$ to $n$}
    \STATE Group each lesion image according to class label.
    \STATE Resize each image to $256\times 256\times 3$.
    \ENDFOR
    \STATE Balance the inter-class data samples.
    \STATE Train CycleGAN on unpaired and balanced interclass data.
    \FOR{$i=1$ to $n$}
    \IF{class label benign}
        \STATE Translate to malignant using the generator network
    \ELSE
        \STATE pass
    \ENDIF
    \ENDFOR
    \STATE Merge synthetic under-represented class outputs and original training set.
    \STATE Shuffle.
    \STATE Train VGG-GAP on the balanced training set
    \STATE Evaluate the model on the test set and generate predicted class labels.
  \end{algorithmic}
\end{algorithm}

\section{Experiments} \label{Experiments}
In this section, extensive experiments are conducted to evaluate the performance of the proposed two-stage approach on a standard benchmark dataset for skin lesion analysis.

\begin{figure}[!htb]
\setlength{\tabcolsep}{.25em}
\centering
\begin{tabular}{cc}
\includegraphics[scale=0.11]{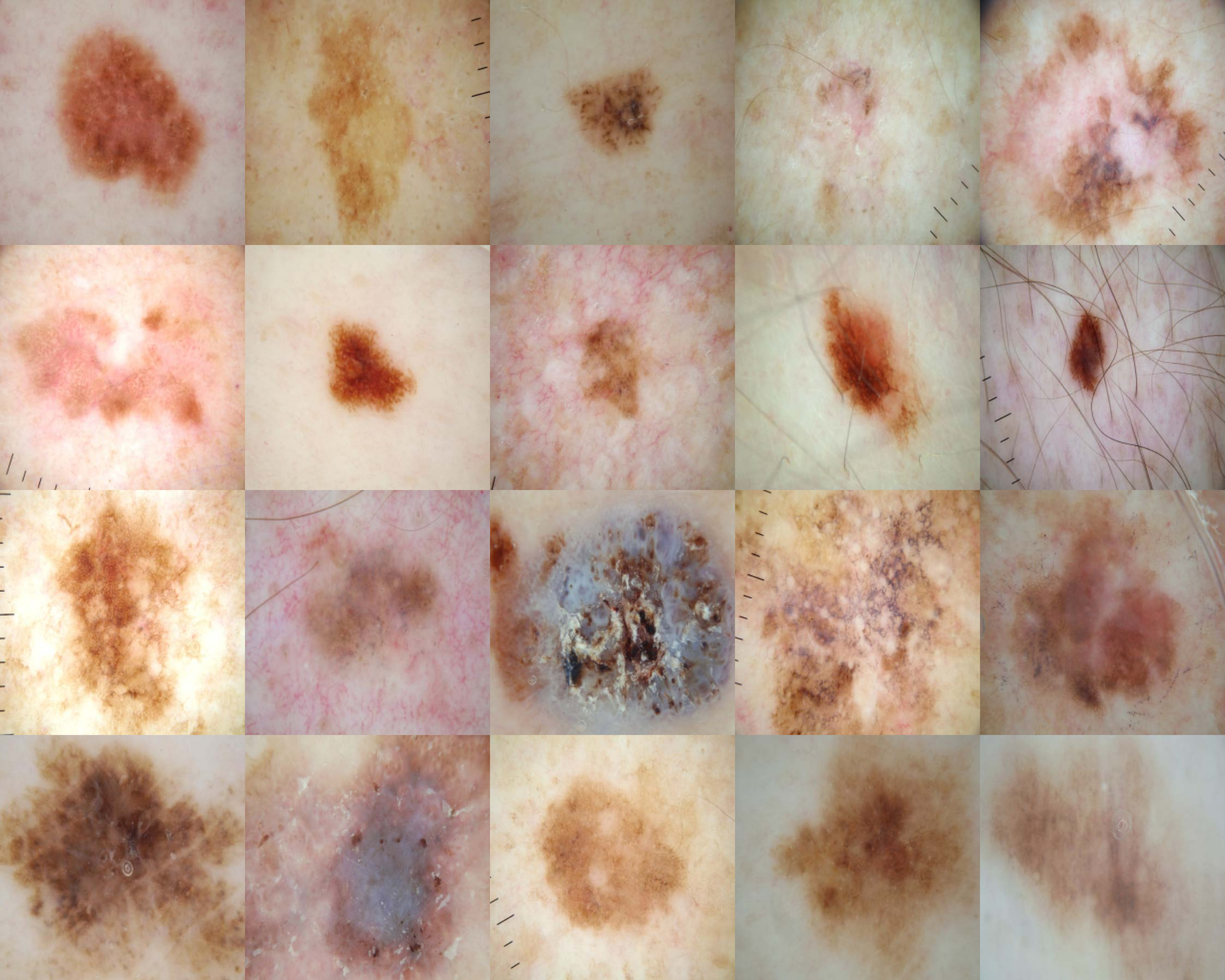} &
\includegraphics[scale=0.11]{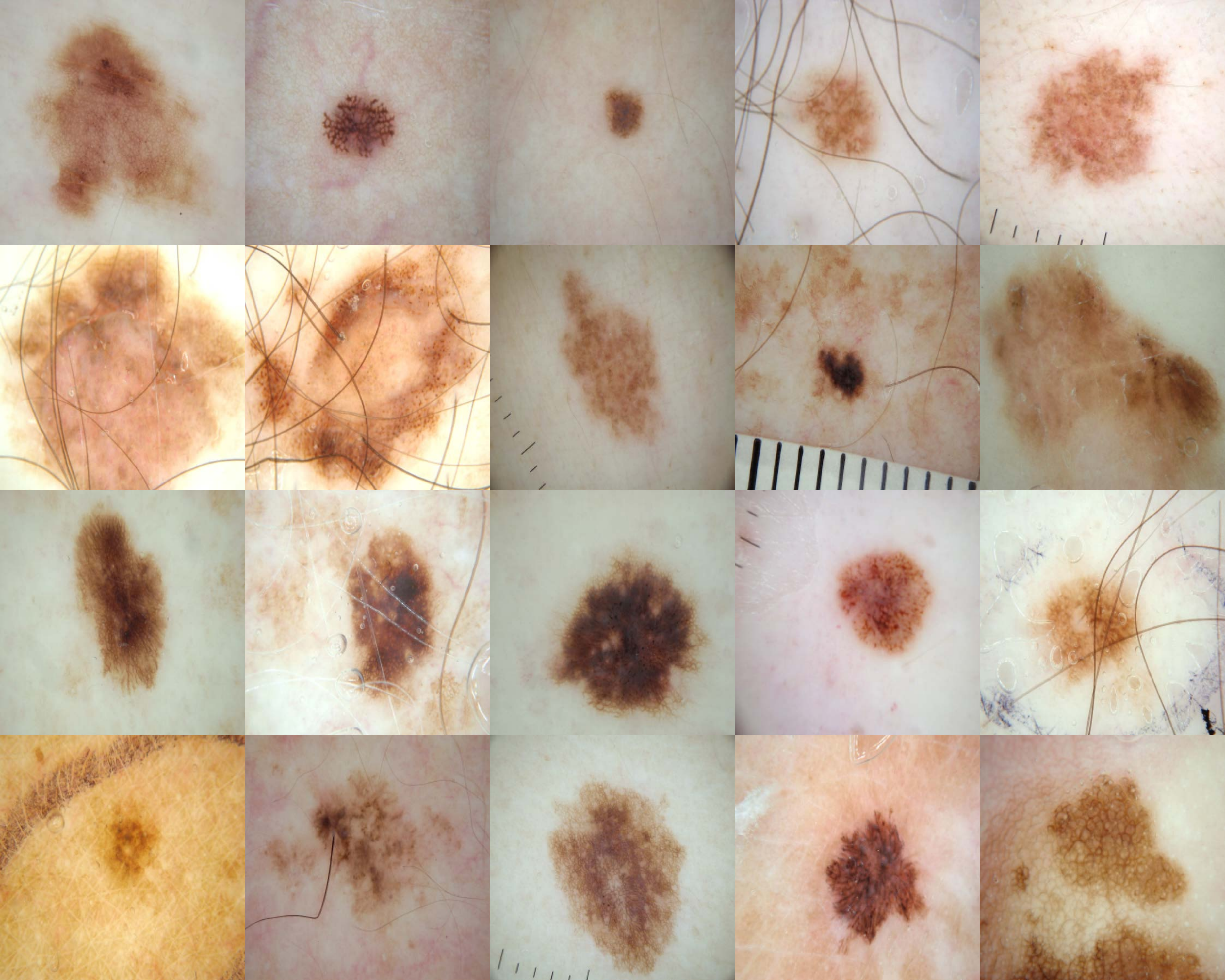}
\end{tabular}
\caption{Sample malignant and benign images from the ISIC-2016 dataset. Notice a high intra-class variation among the malignant samples (left), while benign samples (right) are not visually very different.}
\label{Fig:cancer}
\end{figure}

\medskip
\noindent{\textbf{Dataset.}}\quad \label{Dataset} The effectiveness of MelaNet is evaluated on the ISIC-2016 dataset, a publicly accessible dermatology image analysis benchmark challenge for skin lesion analysis towards melanoma detection~\cite{codella2018skin}, which leverages annotated skin lesion images from the International Skin Imaging Collaboration (ISIC) archive. The dataset contains a representative mix of images of both malignant and benign skin lesions, which were randomly partitioned into training and test sets, with 900 images in the training set and 379 images in the test set, respectively. These images consist of different types of textures in both background and foreground, and also have poor contrast, making the task of melanoma detection a challenging problem. It is also noteworthy to mention that in the training set, there are 727 benign cases and only 173 malignant cases, resulting in an inter-class ratio of 1:4. Sample benign and malignant images from the ISIC-2016 dataset are depicted in Figure~\ref{Fig:cancer}, which shows that both categories have a high visual similarity, making the task of melanoma detection quite arduous. Note that there is a high intra-class variation among the malignant samples. These variations include color, texture and shape. On the other hand, it is important to point out that benign samples are not visually very different, and hence they exhibit low inter-class variation. Furthermore, there are artifacts present in the images such as ruler markers and fine hair, which cause occlusions. Notice that most malignant images show more diffuse boundaries owing to the possibility that before image acquisition, the patient was already diagnosed with melanoma and the medical personnel acquired the dermoscopic images at a deeper level in order to better differentiate between the benign and malignant classes.

\begin{figure}[!htb]
\centering
\includegraphics[scale=.47]{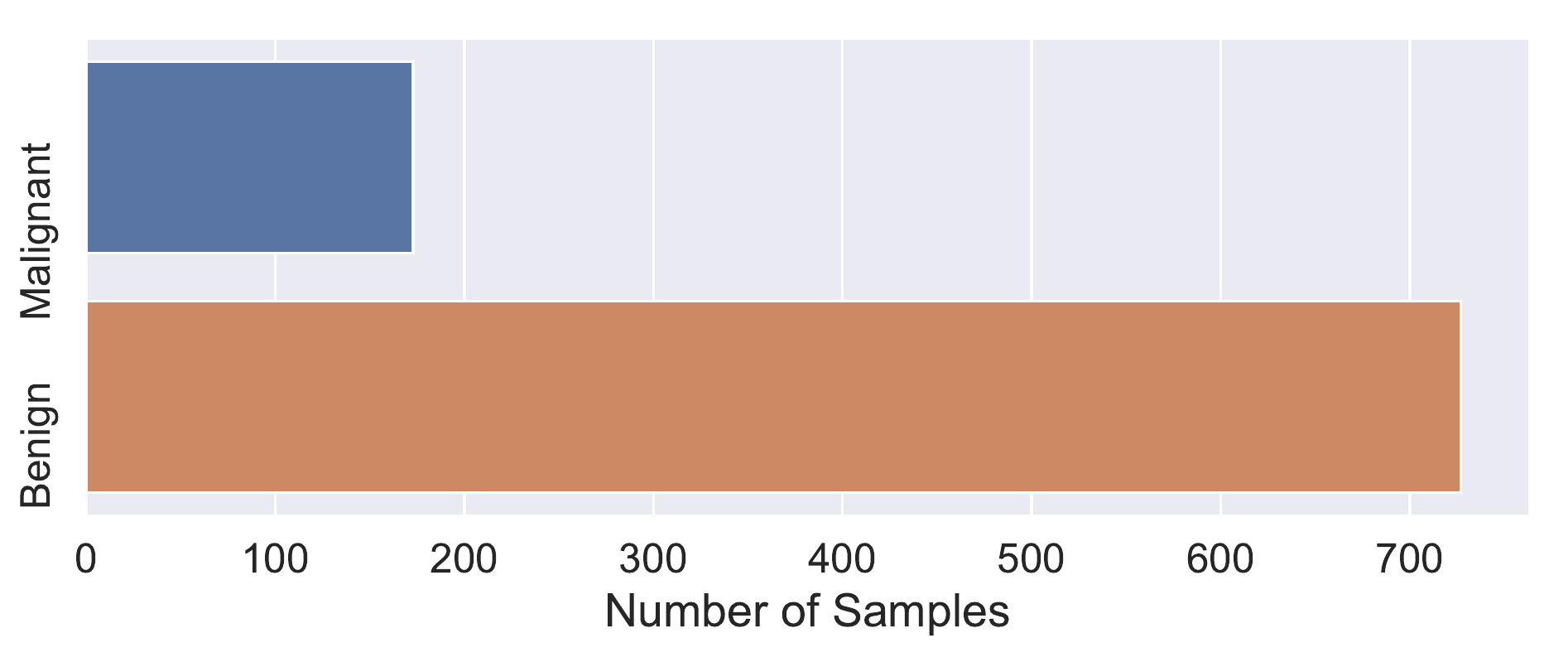}
\caption{Histogram of the ISIC-2016 training set, showing the class imbalance between malignant and benign cases.}
\label{Fig:hist}
\end{figure}

The histogram of the training data is displayed in Figure~\ref{Fig:hist}, showing the class imbalance problem, where the number of images belonging to the minority class (``malignant'') is far smaller than the number of the images belonging to the majority class (``benign''). Also, the number of benign and malignant cases in the test set are 304 and 75, respectively, with an inter-class ratio of 1:4.

Since the images in the ISIC-2016 dataset are of varying sizes, we resize them to $256\times 256$ pixels after applying padding to make them square in order to retain the original aspect ratio.

\medskip
\noindent{\textbf{Training Procedure.}}\quad Since we are tackling a binary classification problem with imbalanced data, we use the focal loss function for the training of the VGG-GAP model. The focal loss is designed to address class imbalance problem by down-weighting easy examples, and focusing more on training the hard examples. Fine-tuning is essentially performed through re-training the whole VGG-GAP network by iteratively minimizing the focal loss function.

\medskip
\noindent{\textbf{Baseline methods.}}\quad  We compare the proposed MelaNet approach against VGG-GAP, VGG-GAP + Augment-5x, and VGG-GAP + Augment-10x. The VGG-GAP network is trained on the original training set, which consists of 900 samples. The VGG-GAP + Augment-5x model uses the same VGG-GAP architecture, but is trained on an augmented dataset composed of 5400 training samples, i.e. we increase the training set 5 times from 900 to 5400 samples using image augmentation. Similarly, the VGG-GAP + Augment-10x network is trained on an augmented set of 99000 training samples (i.e. 10 times the original set). We also ran experiments with augmented training sets higher than 10x the original one, but we did not observe improved performance as the network tends to learn redundant representations.

\medskip
\noindent{\textbf{Implementation details.}}\quad All experiments are carried out on a Linux server with 2x Intel Xeon E5-2650 V4 Broadwell @ 2.2GHz, 256 GB RAM, 4x NVIDIA P100 Pascal (12G HBM2 memory) GPU cards. The algorithms are implemented in Keras with TensorFlow backend.

We train CycleGAN for 500 epochs using Adam optimizer~\cite{kingma2014adam} with learning rate 0.0002 and batch size 1. We set the regularization parameter $\lambda$ to 10. The VGG-GAP classifier, on the other hand, is trained using Adadelta optimizer~\cite{zeiler2012adadelta} with learning rate 0.001 and mini-batch 16. A factor of 0.1 is used to reduce the learning rate once the loss stagnates. For the VGG-GAP model, we set the focal loss parameters to $\alpha=0.25$ and $\gamma=2$, meaning that $\alpha_{t}=0.25$ for positive labeled samples, and $\alpha_{t}=0.75$ for negative labeled samples. Training of VGG-GAP is continued on all network layers until the focal loss stops improving, and then the best weights are retained. For fair comparison, use used the same set of hyper-parameters for VGG-GAP and baseline methods. We choose Adadelta as an optimizer due to its robustness to noisy gradient information and minimal computational overhead.

\subsection{Results}
The effectiveness of the proposed classifier is assessed by conducting a comprehensive comparison with the baseline methods using several performance evaluation metrics~\cite{gutman2016skin,yu2016automated,perez2018data}, including the receiver operating characteristic (ROC) curve, sensitivity, and the area under the ROC curve (AUC). Sensitivity is defined as the percentage of positive instances correctly classified, i.e.
\begin{equation}
\text{Sensitivity} = \frac{\text{TP}}{\text{TP}+\text{FN}},
\label{eq:sens}
\end{equation}
where TP, FP, TN and FN denote true positives, false positives, true negatives and false negatives, respectively. TP is the number of correctly predicted malignant lesions, while TN is the number of correctly predicted benign lesions. A classifier that reduces FN (ruling cancer out in cases that do have it) and FP (wrongly diagnosing cancer where there is none) indicates a better performance. Sensitivity, also known as recall or true positive rate (TPR), indicates how often a classifier misses a positive prediction. It is one of the most common measures to evaluate a classifier in  medical image classification tasks~\cite{esteva2017dermatologist}. We use a threshold of 0.5.

Another common metric is AUC that summarizes the information contained in the ROC curve, which plots TPR versus $\text{FPR}=\text{FP}/(\text{FP}+\text{TN})$, the false positive rate, at various thresholds. Larger AUC values indicate better performance at distinguishing between melonoma and non-melanoma images. It is worth pointing out that the accuracy metric is not used in this study, as it provides no interpretable information and may lead to a false sense of superiority of classifying the majority class.

\begin{figure}[!htb]
\centering
\includegraphics[scale=.62]{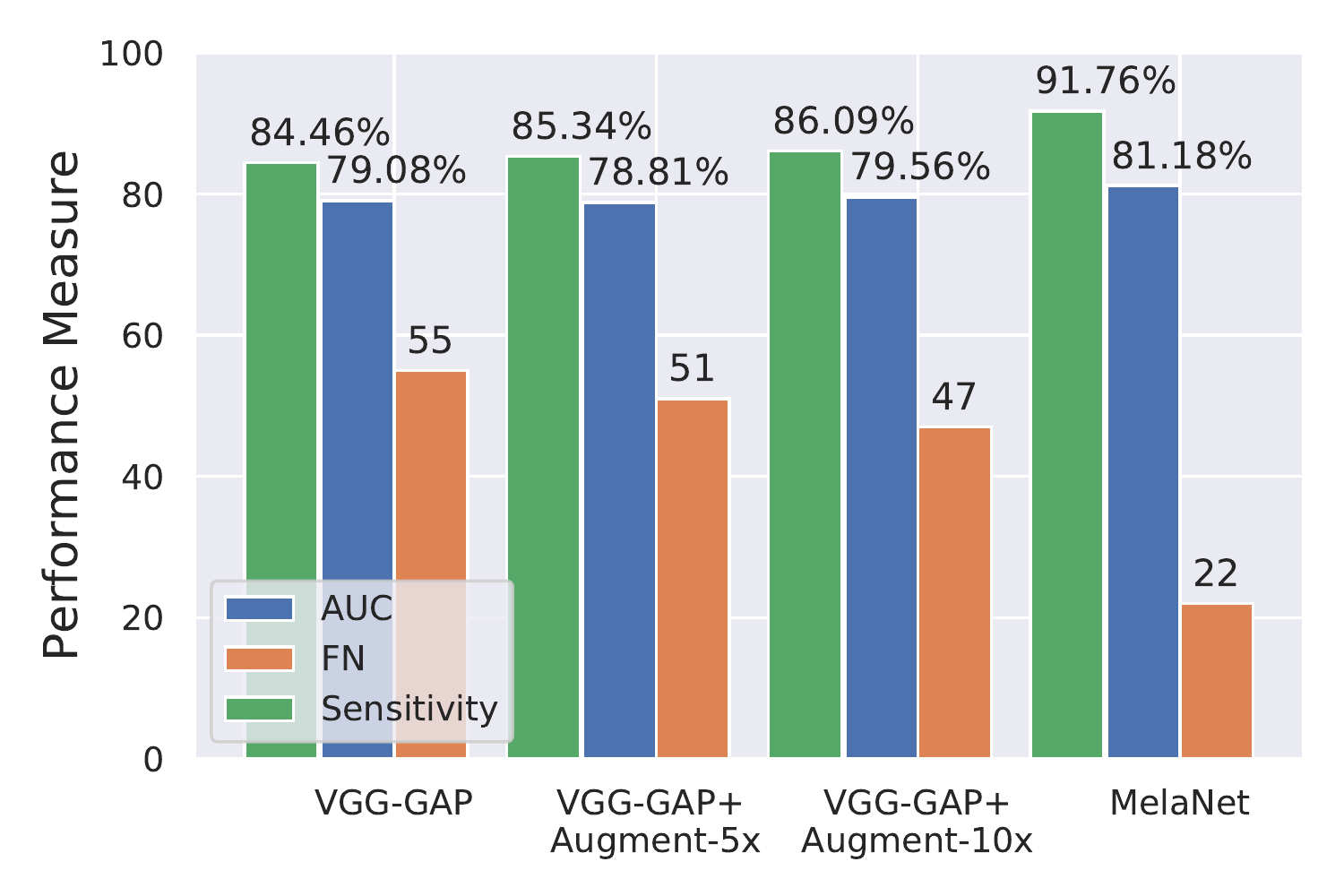}
\caption{Classification performance of MelaNet and the baseline methods using AUC, FN and Sensitivity as evaluation metrics on the ISIC-2016 test set.}
\label{Fig:metrics}
\end{figure}

The performance comparison results of MelaNet and the baseline methods using AUC, FN and Sensitivity are depicted in Figure~\ref{Fig:metrics}. We observe that our approach outperforms the baselines, achieving an AUC of 81.18\% and a sensitivity of 91.76\% with performance improvements of 2.1\% and 7.3\% over the VGG-GAP baseline. Interestingly, MelaNet yields the lowest number of false negatives, which were reduced by more than 50\% compared to the baseline methods, meaning it picked up on malignant cases that the baselines had missed. In other words, MelaNet caught instances of melanoma that would have otherwise gone undetected. This is a significant performance in the potential for early melanoma detection, albeit MelaNet was trained on only 1627 samples composed of 900 images from the original dataset and 727 synthesized images (benign and malignant) obtained via generative adversarial training.

\begin{figure}[!htb]
\centering
\includegraphics[scale=.65]{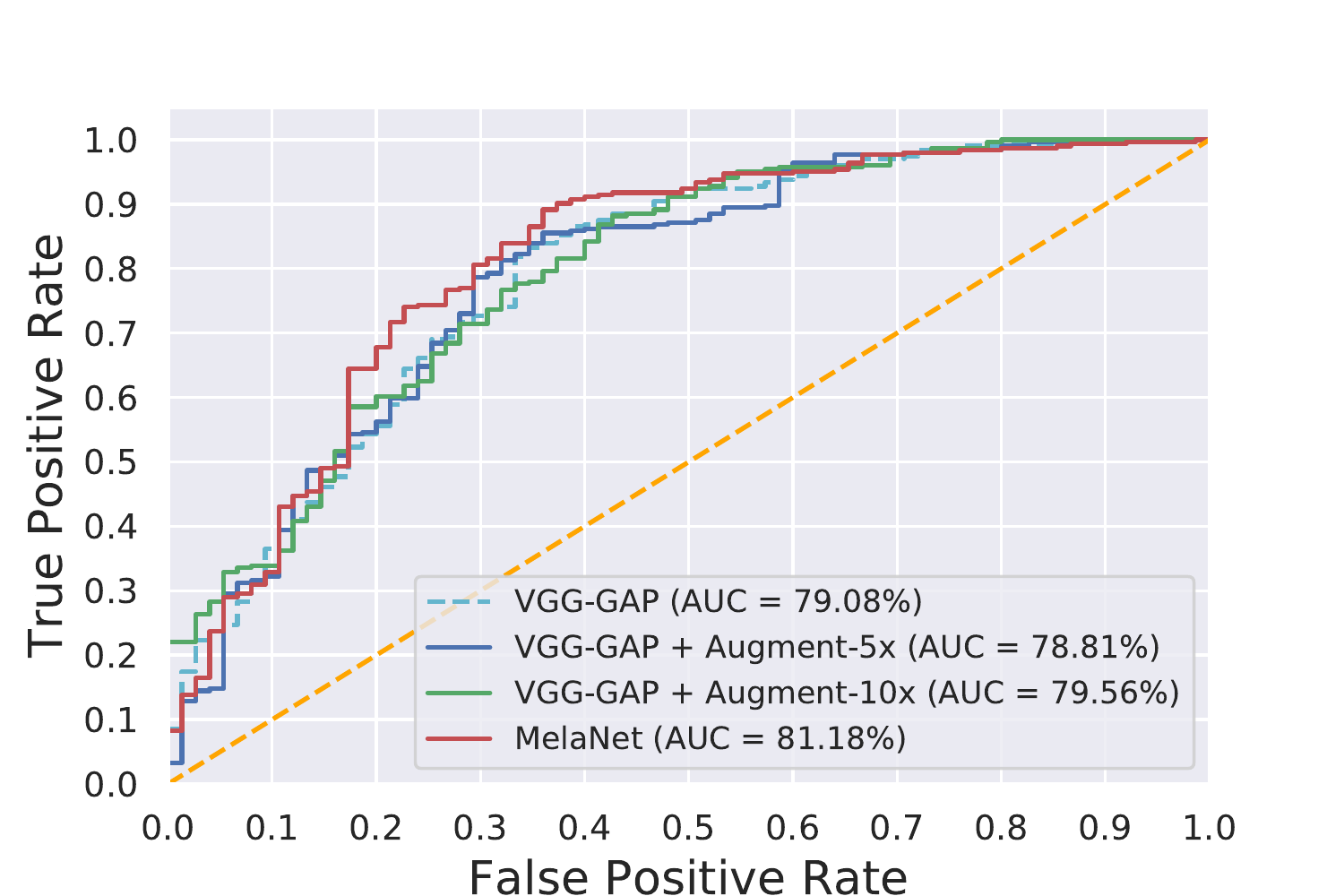}
\caption{ROC curves for MelaNet and baseline methods, along with the corresponding AUC values.}
\label{Fig:auc}
\end{figure}

Figure~\ref{Fig:auc} displays the ROC curve, which shows the better performance of our proposed MelaNet approach compared to the baseline methods. Each point on ROC represents different trade-off between false positives and false negatives. An ROC curve that is closer to the upper right indicates a better performance (TPR
is higher than FPR). Even though during the early and last stages, the ROC curve of MelaNet seems to fluctuate at certain points, the overall performance is much higher than the baselines, as indicated by the AUC value. This better performance demonstrates that the conditional image synthesis procedure plays a crucial role and enables our model to learn effective representations, while mitigating data scarcity and class imbalance.

We also compare MelaNet to two other standard baseline methods~\cite{gutman2016skin,yu2016automated}. The top evaluation results on the ISIC-2016 dataset to classify images as either being benign or malignant are reported in~\cite{gutman2016skin}. The method presented in~\cite{yu2016automated} is also a two-stage approach consisting of a fully convolutional residual network for skin lesion segmentation, followed by a very deep residual network for skin lesion classification. The classification results are displayed in Table~\ref{Tab:class}, which shows that the proposed approach achieves significantly better results than the baseline methods.

\begin{table*}[!tb]
\rowcolors{1}{}{lightblue}
\caption{Classification evaluation results of MelaNet and baseline methods. Boldface numbers indicate the best performance.}
\medskip
\centering
\begin{tabular}{l*{3}{c}}
\toprule
& \multicolumn{3}{c}{Performance Measures}\\
\cmidrule(lr){2-4}
Method &  AUC (\%) & Sensitivity (\%) & FN\\
\midrule
Gutman \textit{et al.}~\cite{gutman2016skin}~ & 80.40 & 50.70 & -- \\
Yu \textit{et al.}~\cite{yu2016automated} (\textit{without segmentation})~ & 78.20 & 42.70 & -- \\
Yu \textit{et al.}~\cite{yu2016automated} (\textit{with segmentation})~ & 78.30 & 54.70 & -- \\
VGG-GAP  ~ & 79.08 & 84.46 & 55 \\
VGG-GAP + Augment-5x (\textit{ours}) ~ & 78.81 & 85.34 & 51 \\
VGG-GAP + Augment-10x (\textit{ours}) ~ & 79.56 & 86.09 & 47 \\
MelaNet (\textit{ours}) ~ & \textbf{81.18} & \textbf{91.76} & \textbf{22} \\
\bottomrule
\end{tabular}
\label{Tab:class}
\end{table*}

\medskip
\noindent{\textbf{Feature visualization and analysis.}}\quad Understanding and interpreting the predictions made by a deep learning model provides valuabe insights into the input data and the features learned by the model so that the results can be easily understood by human experts. In order to visually explain the decisions made by the proposed classifier and baseline methods, we use gradient-weighted class activation map (Grad-CAM)~\cite{selvaraju2017grad} to generate the saliency maps that highlight the most influential features affecting the predictions. Since convolutional feature maps retain spatial information and each pixel of the feature map indicates whether the corresponding visual pattern exists in its receptive field, the output from the last convolutional layer of the VGG-16 network shows the discriminative region of the image.

\begin{figure}[!htb]
\centering
\includegraphics[scale=.77]{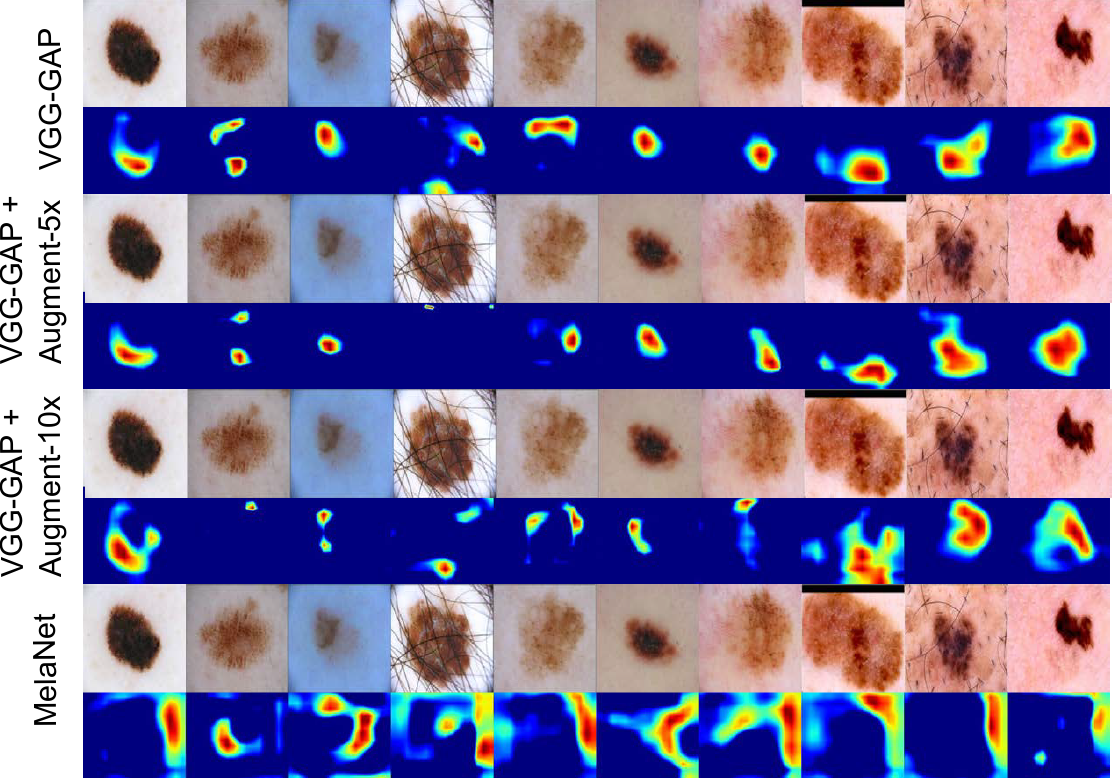}
\caption{Grad-CAM heat maps for the misclassified malignant cases by MelaNet and baseline methods.}
\label{Fig:miss}
\end{figure}

The class activation maps displayed in Figure~\ref{Fig:miss} show that even though the baseline methods demonstrate high activations for the region consisting of the lesion, they still fail to correctly classify the dermoscopic image. For our proposed MelaNet approach, we observe that the area surrounding the skin lesion is highly activated. Notice that most of the borders of the whole input image are highlighted, due largely to the fact the classifiers are not looking at the regions of interest, and hence result in misclassification.

\begin{figure}[!htb]
\centering
\includegraphics[scale=.77]{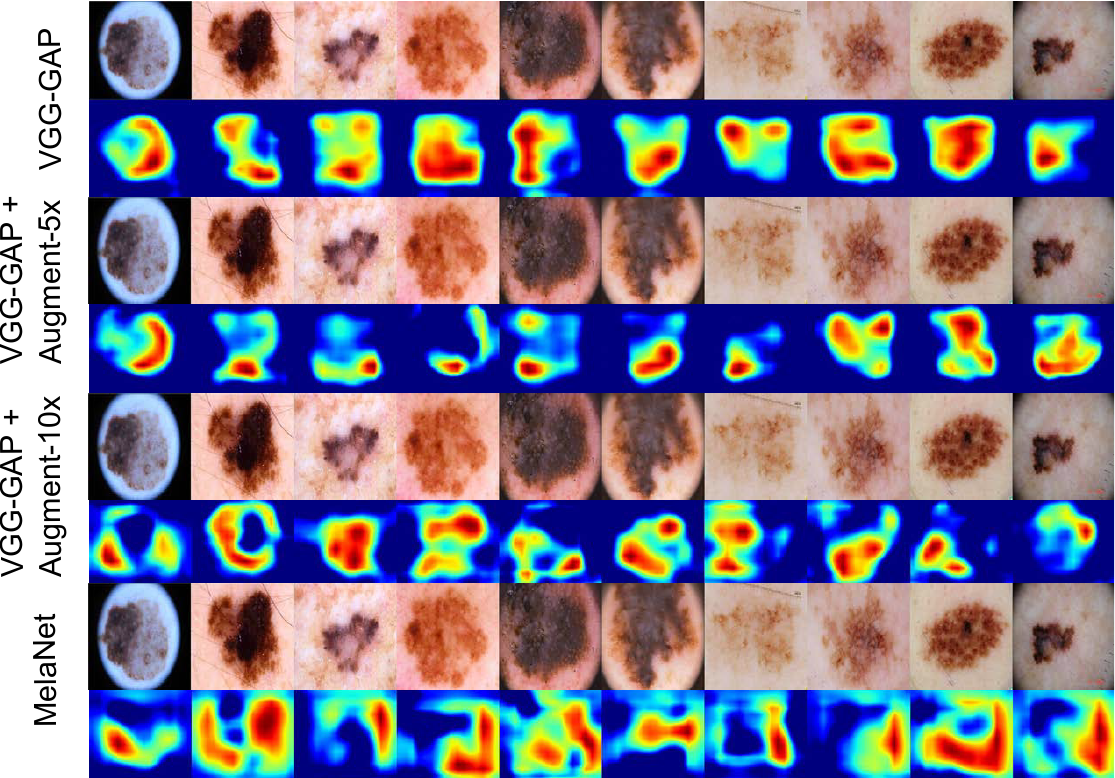}
\caption{Grad-CAM heat maps for the correctly classified malignant cases by MelaNet and baseline methods.}
\label{Fig:true}
\end{figure}

We can also see in Figure~\ref{Fig:true} that while the proposed approach shows similar visual patterns as the baselines when correctly classifying the input image, it, however, outputs high activations for the regions surrounding the skin lesion in many cases. These regions consist of shapes and edges. Hence, our approach not only focus on the skin lesion, but also captures its context, which helps in the final detection. This context-based approach is commonly used by expert dermatologists~\cite{esteva2017dermatologist}. This observation is of great significance, and further shows the effectiveness of our approach.

\begin{figure}[!htb]
\setlength{\tabcolsep}{.1em}
\centering
 \begin{tabular}{cc}
\fbox{\includegraphics[width=1.6in,height=1.5in]{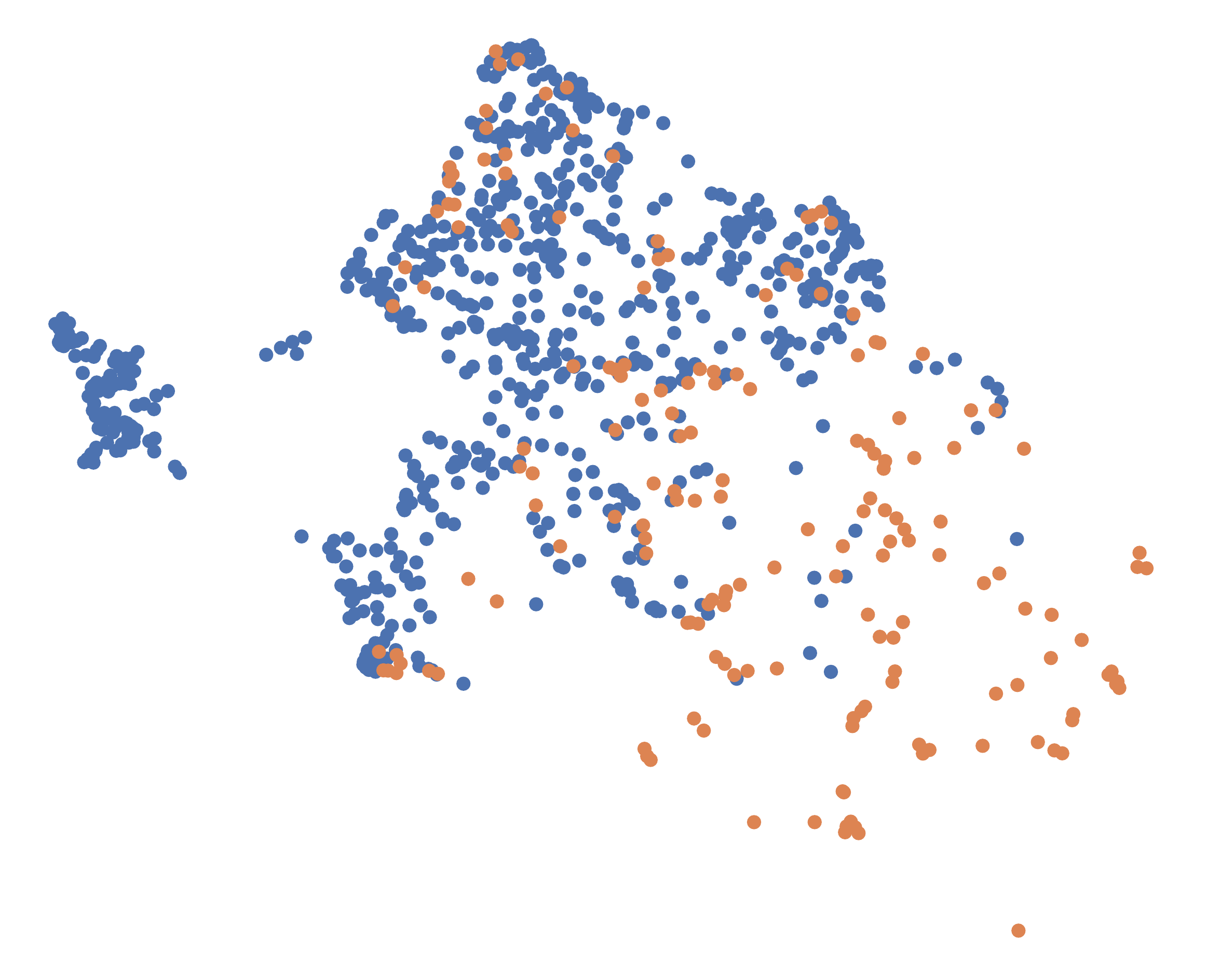}} &
\fbox{\includegraphics[width=1.6in,height=1.5in]{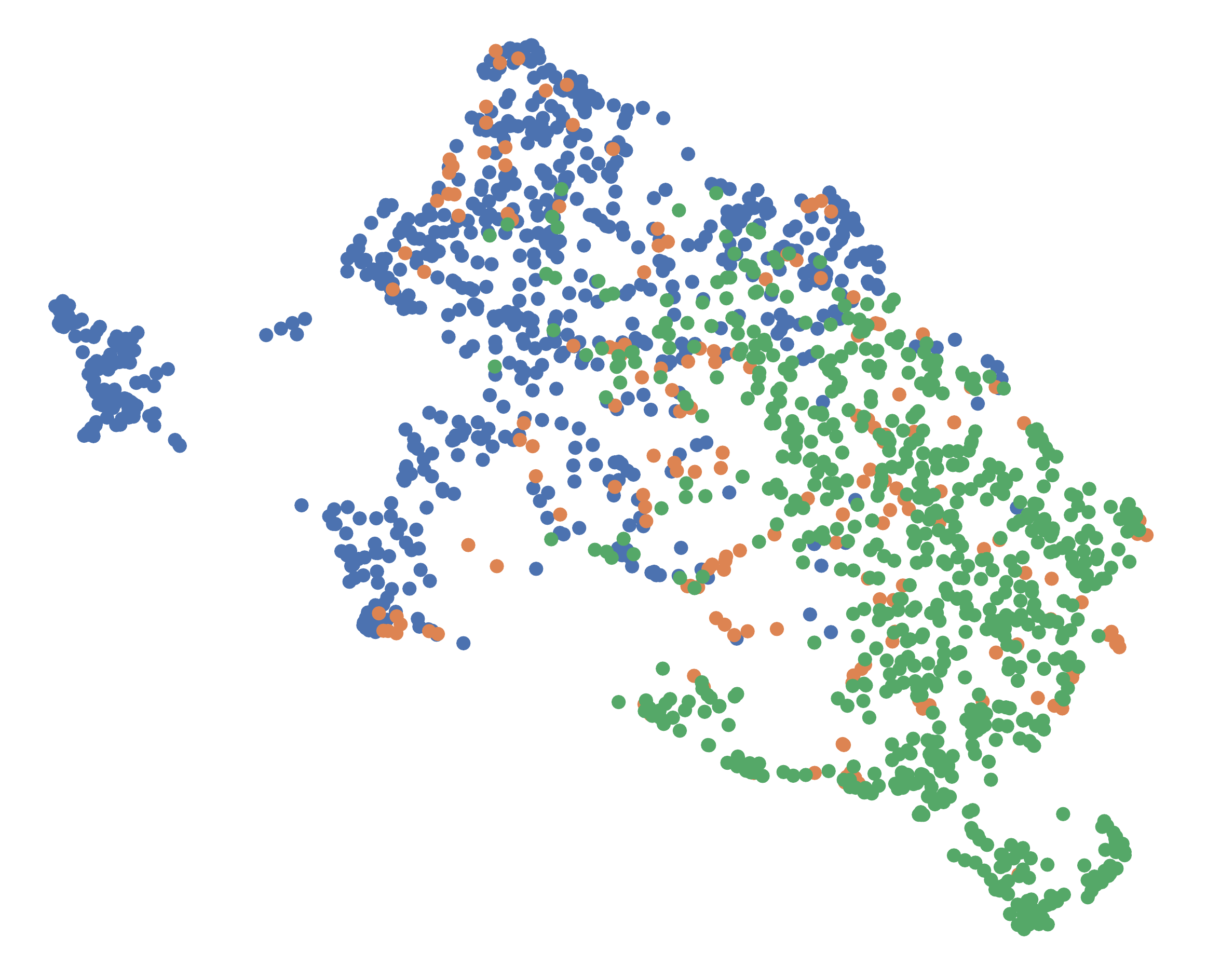}}
\end{tabular}
\caption{Two-dimensional UMAP embeddings using the original ISIC-2016 training set (left) consisting of 900 samples (benign shown in blue and malignant in orange) and with additional synthesized malignant data samples (shown in green) consisting of a total 1627 samples (right).}
\label{Fig:UMAP}
\end{figure}

In order to get a clear understanding of the data distribution, the learned features from both the original training set and the balanced dataset (i.e. with the additional synthesized data using adversarial training) are visualized using Uniform Manifold Approximation and Projection for Dimension Reduction (UMAP)~\cite{mcinnes2018umap}, which is a dimensionality reduction technique that is particularly well-suited for embedding high-dimensional data into a two- or three-dimensional space. The UMAP embeddings shown in Figure \ref{Fig:UMAP} were generated by running the UMAP algorithm on the original training set with 900 samples (benign and malignant) and the balanced dataset consisting of 1627 samples (benign and malignant).

From Figure~\ref{Fig:UMAP} (left), it is evident that the inter-class variation is significantly small due in large part to the very high visual similarity between malignant and benign skin lesions. Hence, the task of learning a decision boundary between the two categories is challenging. We can also see that the synthesized samples (malignant lesions shown in green) lie very close to the original data distribution. It is important to note that the outliers present in the dataset are not due to the image synthesis procedure, but this is rather a characteristic present in the original training set. Therefore, the synthetically generated data are representative of the original under-represented class (i.e. malignant skin lesions).

\begin{figure}[!htb]
\centering
\includegraphics[scale=.35]{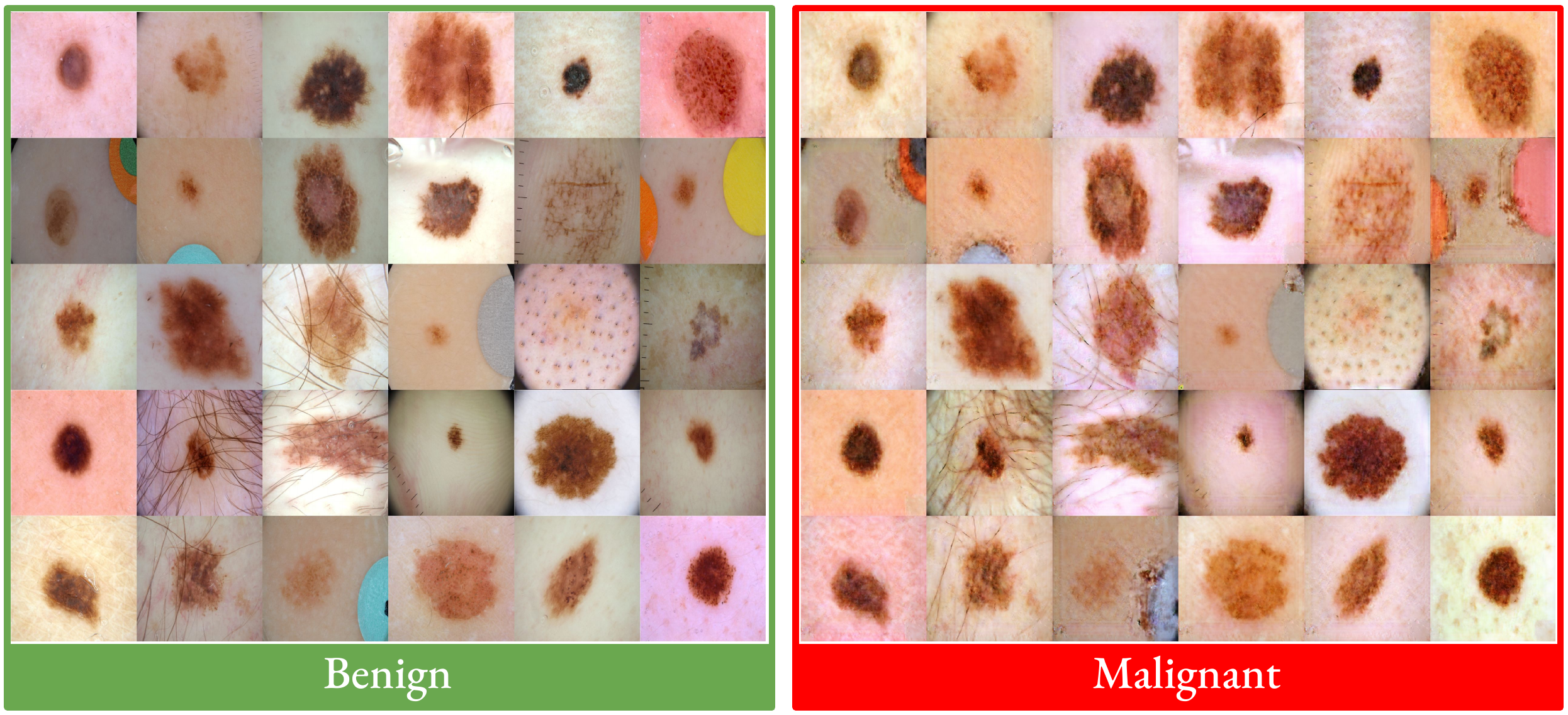}
\caption{Sample benign images from the ISIC-2016 dataset that are translated to malignant images using the proposed approach. Notice that the synthesized images display a reasonably good visual quality. }
\label{Fig:gangan}
\end{figure}

\medskip
\noindent{\textbf{Discussion.}}\quad \label{Discussion} With a training set consisting of only 1627 images, our proposed MelaNet approach is able to achieve improved performance. This better performance is largely due to the fact that by leveraging the inter-class variation in medical images, the mapping between the source and target distribution for conditional image synthesis can be easily learned. Moreover, it is much easier to generate target images given prior information, rather than generating from noise which often results in training instability and artifacts~\cite{perez2018data}. It is important to note that even though image-to-image translation schemes are considered to hallucinate images by adding or removing image features~\cite{cohen2018distribution}, we showed that in our scheme the partition of the inter-classes does not result in a bias or unwanted feature hallucination. Figure~\ref{Fig:gangan} shows the benign lesions sampled from the ISIC-2016 training set, which are translated to malignant samples using MelaNet. As can be seen, the benign and the corresponding synthesized malignant images have a high degree of visual similarity. This is largely due to the nature of the dataset, which is known to have a low inter-class variation.

In the synthetic minority over-sampling technique (SMOTE), when drawing random observations from its k-nearest neighbors, it is possible that a ``border point'' or an observation very close to the decision boundary may be selected, resulting in synthetically-generated observations lying too close to the decision boundary, and as a result the performance of the classifier may be degraded. The advantage of our approach over SMOTE is that we learn a transformation between a source and a target domain by solving an optimization problem in order to determine two bijective mappings. This enables the generator to synthesize observations, which help improve the classification performance while learning the transformation/decision boundary.

In order to gain a deeper insight on the performance of the proposed approach, we sample all the original benign lesions and a subset of the synthesized malignant lesions, consisting of 727 and 10 samples, respectively. For the benign group of images, the proposed MelaNet model yields a sensitivity score of 89\%, with 77 misclassified images. By contrast, a 100\% sensitivity score is obtained when performing predictions on the synthesized malignant group of images. In addition, the F-score values for MelaNet on the benign and synthesized malignant groups are 94\% and 21\%, respectively.
\section{Conclusion}
In this paper, we proposed a two-stage framework for melanoma detection. The first stage addresses the problem of data scarcity and class imbalance by formulating inter-class variation as conditional image synthesis for over-sampling in order to synthesize under-represented class samples (e.g. melanoma from non-melanoma lesions). The newly synthesized samples are then used as additional data to train a deep convolutional neural network by minimizing the focal loss function, which assists the classifier in learning from hard examples. We demonstrate through extensive experiments that the proposed MelaNet approach improves sensitivity by a margin of 13.10\% and the AUC by 0.78\% from only 1627 dermoscopy images compared to the baseline methods on the ISIC-2016 dataset. For future work directions, we plan to address the multi-class classification problem, which requires an independent generative model for each domain, leading to prohibitive computational overhead for adversarial training. We also intend to apply our method to other medical imaging modalities.

\bibliographystyle{ieeetr}
\bibliography{references}

\begin{thebibliography}{10}

\bibitem{Saito:2018}
E.~Saito and M.~Hori, ``Melanoma skin cancer incidence rates in the world from
  the cancer incidence in five continents {XI},'' {\em Japanese Journal of
  Clinical Oncology}, vol.~48, no.~12, pp.~1113--1114, 2018.

\bibitem{Siegel:2019}
R.~Siegel, K.~Miller, , and A.~Jemal, ``Cancer statistics, 2019,'' {\em CA: A
  Cancer Journal for Clinicians}, vol.~69, pp.~7--34, 2019.

\bibitem{binder1995epiluminescence}
M.~Binder, M.~Schwarz, A.~Winkler, A.~Steiner, A.~Kaider, K.~Wolff, and
  H.~Pehamberger, ``Epiluminescence microscopy: a useful tool for the diagnosis
  of pigmented skin lesions for formally trained dermatologists,'' {\em
  Archives of Dermatology}, vol.~131, no.~3, pp.~286--291, 1995.

\bibitem{ganster2001automated}
H.~Ganster, P.~Pinz, R.~Rohrer, E.~Wildling, M.~Binder, and H.~Kittler,
  ``Automated melanoma recognition,'' {\em IEEE Transactions on Medical
  Imaging}, vol.~20, no.~3, pp.~233--239, 2001.

\bibitem{cheng2008skin}
Y.~Cheng, R.~Swamisai, S.~E. Umbaugh, R.~H. Moss, W.~V. Stoecker, S.~Teegala,
  and S.~K. Srinivasan, ``Skin lesion classification using relative color
  features,'' {\em Skin Research and Technology}, vol.~14, no.~1, pp.~53--64,
  2008.

\bibitem{ZLiu:15}
Z.~Liu and J.~Zerubia, ``Skin image illumination modeling and chromophore
  identification for melanoma diagnosis,'' {\em Physics in Medicine \&
  Biology}, vol.~60, pp.~3415--3431, 2015.

\bibitem{mishra2016overview}
N.~K. Mishra and M.~E. Celebi, ``An overview of melanoma detection in
  dermoscopy images using image processing and machine learning,'' {\em arXiv
  preprint arXiv:1601.07843}, 2016.

\bibitem{ballerini2013color}
L.~Ballerini, R.~B. Fisher, B.~Aldridge, and J.~Rees, ``A color and texture
  based hierarchical {K-NN} approach to the classification of non-melanoma skin
  lesions,'' in {\em Color Medical Image Analysis}, pp.~63--86, Springer, 2013.

\bibitem{tommasi2006melanoma}
T.~Tommasi, E.~La~Torre, and B.~Caputo, ``Melanoma recognition using
  representative and discriminative kernel classifiers,'' in {\em Proc.
  International Workshop on Computer Vision Approaches to Medical Image
  Analysis}, pp.~1--12, 2006.

\bibitem{celebi2007methodological}
M.~E. Celebi, H.~A. Kingravi, B.~Uddin, H.~Iyatomi, Y.~A. Aslandogan, W.~V.
  Stoecker, and R.~H. Moss, ``A methodological approach to the classification
  of dermoscopy images,'' {\em Computerized Medical Imaging and Graphics},
  vol.~31, no.~6, pp.~362--373, 2007.

\bibitem{schaefer2014ensemble}
G.~Schaefer, B.~Krawczyk, M.~E. Celebi, and H.~Iyatomi, ``An ensemble
  classification approach for melanoma diagnosis,'' {\em Memetic Computing},
  vol.~6, no.~4, pp.~233--240, 2014.

\bibitem{krizhevsky2012imagenet}
A.~Krizhevsky, I.~Sutskever, and G.~E. Hinton, ``{ImageNet} classification with
  deep convolutional neural networks,'' in {\em Advances in Neural Information
  Processing Systems}, pp.~1097--1105, 2012.

\bibitem{ronneberger2015u}
O.~Ronneberger, P.~Fischer, and T.~Brox, ``{U-Net}: Convolutional networks for
  biomedical image segmentation,'' in {\em Proc. International Conference on
  Medical Image Computing and Computer-Assisted Intervention}, pp.~234--241,
  2015.

\bibitem{roth2014new}
H.~R. Roth, L.~Lu, A.~Seff, K.~M. Cherry, J.~Hoffman, S.~Wang, J.~Liu,
  E.~Turkbey, and R.~M. Summers, ``A new 2.5 d representation for lymph node
  detection using random sets of deep convolutional neural network
  observations,'' in {\em Proc. International Conference on Medical Image
  Computing and Computer-Assisted Intervention}, pp.~520--527, 2014.

\bibitem{anthimopoulos2016lung}
M.~Anthimopoulos, S.~Christodoulidis, L.~Ebner, A.~Christe, and S.~Mougiakakou,
  ``Lung pattern classification for interstitial lung diseases using a deep
  convolutional neural network,'' {\em IEEE Transactions on Medical Imaging},
  vol.~35, no.~5, pp.~1207--1216, 2016.

\bibitem{matsunaga2017image}
K.~Matsunaga, A.~Hamada, A.~Minagawa, and H.~Koga, ``Image classification of
  melanoma, nevus and seborrheic keratosis by deep neural network ensemble,''
  {\em arXiv preprint arXiv:1703.03108}, 2017.

\bibitem{codella2017deep}
N.~C. Codella, Q.-B. Nguyen, S.~Pankanti, D.~A. Gutman, B.~Helba, A.~C.
  Halpern, and J.~R. Smith, ``Deep learning ensembles for melanoma recognition
  in dermoscopy images,'' {\em IBM Journal of Research and Development},
  vol.~61, no.~4/5, pp.~5--1, 2017.

\bibitem{gutman2016skin}
D.~Gutman, N.~C. Codella, E.~Celebi, B.~Helba, M.~Marchetti, N.~Mishra, and
  A.~Halpern, ``Skin lesion analysis toward melanoma detection: A challenge at
  the international symposium on biomedical imaging ({ISBI}) 2016, hosted by
  the international skin imaging collaboration ({ISIC}),'' {\em arXiv preprint
  arXiv:1605.01397}, 2016.

\bibitem{yu2016automated}
L.~Yu, H.~Chen, Q.~Dou, J.~Qin, and P.-A. Heng, ``Automated melanoma
  recognition in dermoscopy images via very deep residual networks,'' {\em IEEE
  Transactions on Medical Imaging}, vol.~36, no.~4, pp.~994--1004, 2016.

\bibitem{shie2015transfer}
C.-K. Shie, C.-H. Chuang, C.-N. Chou, M.-H. Wu, and E.~Y. Chang, ``Transfer
  representation learning for medical image analysis,'' in {\em Proc.
  International Conference of the IEEE Engineering in Medicine and Biology
  Society}, pp.~711--714, 2015.

\bibitem{shin2016deep}
H.-C. Shin, H.~R. Roth, M.~Gao, L.~Lu, Z.~Xu, I.~Nogues, J.~Yao, D.~Mollura,
  and R.~M. Summers, ``Deep convolutional neural networks for computer-aided
  detection: {CNN} architectures, dataset characteristics and transfer
  learning,'' {\em IEEE Transactions on Medical Imaging}, vol.~35, no.~5,
  pp.~1285--1298, 2016.

\bibitem{goodfellow2014generative}
I.~Goodfellow, J.~Pouget-Abadie, M.~Mirza, B.~Xu, D.~Warde-Farley, S.~Ozair,
  A.~Courville, and Y.~Bengio, ``Generative adversarial nets,'' in {\em
  Advances in Neural Information Processing Systems}, pp.~2672--2680, 2014.

\bibitem{nie2017medical}
D.~Nie, R.~Trullo, J.~Lian, C.~Petitjean, S.~Ruan, Q.~Wang, and D.~Shen,
  ``Medical image synthesis with context-aware generative adversarial
  networks,'' in {\em Proc. International Conference on Medical Image Computing
  and Computer-Assisted Intervention}, pp.~417--425, 2017.

\bibitem{frid2018synthetic}
M.~Frid-Adar, E.~Klang, M.~Amitai, J.~Goldberger, and H.~Greenspan, ``Synthetic
  data augmentation using gan for improved liver lesion classification,'' in
  {\em Proc. IEEE International Symposium on Biomedical Imaging}, pp.~289--293,
  2018.

\bibitem{yi2019generative}
X.~Yi, E.~Walia, and P.~Babyn, ``Generative adversarial network in medical
  imaging: A review,'' {\em Medical Image Analysis}, 2019.

\bibitem{costa2017end}
P.~Costa, A.~Galdran, M.~I. Meyer, M.~Niemeijer, M.~Abr{\`a}moff, A.~M.
  Mendon{\c{c}}a, and A.~Campilho, ``End-to-end adversarial retinal image
  synthesis,'' {\em IEEE Transactions on Medical Imaging}, vol.~37, no.~3,
  pp.~781--791, 2017.

\bibitem{zhang2019skrgan}
T.~Zhang, H.~Fu, Y.~Zhao, J.~Cheng, M.~Guo, Z.~Gu, B.~Yang, Y.~Xiao, S.~Gao,
  and J.~Liu, ``{SkrGAN}: Sketching-rendering unconditional generative
  adversarial networks for medical image synthesis,'' in {\em Proc.
  International Conference on Medical Image Computing and Computer-Assisted
  Intervention}, pp.~777--785, 2019.

\bibitem{zhu2017unpaired}
J.-Y. Zhu, T.~Park, P.~Isola, and A.~A. Efros, ``Unpaired image-to-image
  translation using cycle-consistent adversarial networks,'' in {\em Proc. IEEE
  International Conference on Computer Vision}, pp.~2223--2232, 2017.

\bibitem{isola2017image}
P.~Isola, J.-Y. Zhu, T.~Zhou, and A.~A. Efros, ``Image-to-image translation
  with conditional adversarial networks,'' in {\em Proc. IEEE Conference on
  Computer Vision and Pattern Recognition}, pp.~1125--1134, 2017.

\bibitem{liu2017unsupervised}
M.-Y. Liu, T.~Breuel, and J.~Kautz, ``Unsupervised image-to-image translation
  networks,'' in {\em Advances in Neural Information Processing Systems},
  pp.~700--708, 2017.

\bibitem{lamb2017gibbsnet}
A.~M. Lamb, D.~Hjelm, Y.~Ganin, J.~P. Cohen, A.~C. Courville, and Y.~Bengio,
  ``{GibbsNet}: Iterative adversarial inference for deep graphical models,'' in
  {\em Advances in Neural Information Processing Systems}, pp.~5089--5098,
  2017.

\bibitem{ben2017virtual}
A.~Ben-Cohen, E.~Klang, S.~P. Raskin, M.~M. Amitai, and H.~Greenspan, ``Virtual
  {PET} images from {CT} data using deep convolutional networks: initial
  results,'' in {\em Proc. International Workshop on Simulation and Synthesis
  in Medical Imaging}, pp.~49--57, 2017.

\bibitem{yang2017dagan}
G.~Yang, S.~Yu, H.~Dong, G.~Slabaugh, P.~L. Dragotti, X.~Ye, F.~Liu,
  S.~Arridge, J.~Keegan, Y.~Guo, {\em et~al.}, ``{DAGAN}: deep de-aliasing
  generative adversarial networks for fast compressed sensing mri
  reconstruction,'' {\em IEEE Transactions on Medical Imaging}, vol.~37, no.~6,
  pp.~1310--1321, 2017.

\bibitem{wolterink2017deep}
J.~M. Wolterink, A.~M. Dinkla, M.~H. Savenije, P.~R. Seevinck, C.~A. van~den
  Berg, and I.~I{\v{s}}gum, ``Deep {MR} to {CT} synthesis using unpaired
  data,'' in {\em Proc. International Workshop on Simulation and Synthesis in
  Medical Imaging}, pp.~14--23, 2017.

\bibitem{russ2019synthesis}
T.~Russ, S.~Goerttler, A.-K. Schnurr, D.~F. Bauer, S.~Hatamikia, L.~R. Schad,
  F.~G. Z{\"o}llner, and K.~Chung, ``Synthesis of {CT} images from digital body
  phantoms using {CycleGAN},'' {\em International Journal of Computer Assisted
  Radiology and Surgery}, vol.~14, no.~10, pp.~1741--1750, 2019.

\bibitem{shaban2019staingan}
M.~T. Shaban, C.~Baur, N.~Navab, and S.~Albarqouni, ``{StainGAN}: Stain style
  transfer for digital histological images,'' in {\em Proc. IEEE International
  Symposium on Biomedical Imaging}, pp.~953--956, 2019.

\bibitem{de2018stain}
T.~de~Bel, M.~Hermsen, J.~Kers, J.~van~der Laak, and G.~Litjens,
  ``Stain-transforming cycle-consistent generative adversarial networks for
  improved segmentation of renal histopathology,'' in {\em Proc. International
  Conference on Medical Imaging with Deep Learning}, vol.~102, pp.~151--163,
  2018.

\bibitem{bissoto2018skin}
A.~Bissoto, F.~Perez, E.~Valle, and S.~Avila, ``Skin lesion synthesis with
  generative adversarial networks,'' in {\em Proc. International Workshop on
  Computer-Assisted and Robotic Endoscopy}, pp.~294--302, 2018.

\bibitem{ali2019data}
I.~S. Ali, M.~F. Mohamed, and Y.~B. Mahdy, ``Data augmentation for skin lesion
  using self-attention based progressive generative adversarial network,'' {\em
  arXiv preprint arXiv:1910.11960}, 2019.

\bibitem{cohen2018distribution}
J.~P. Cohen, M.~Luck, and S.~Honari, ``Distribution matching losses can
  hallucinate features in medical image translation,'' in {\em Proc.
  International Conference on Medical Image Computing and Computer-Assisted
  Intervention}, pp.~529--536, 2018.

\bibitem{zhao2019compression}
Z.~Zhao, Q.~Sun, H.~Yang, H.~Qiao, Z.~Wang, and D.~O. Wu, ``Compression
  artifacts reduction by improved generative adversarial networks,'' {\em
  EURASIP Journal on Image and Video Processing}, vol.~2019, no.~1, p.~62,
  2019.

\bibitem{kazeminia2018gans}
S.~Kazeminia, C.~Baur, A.~Kuijper, B.~van Ginneken, N.~Navab, S.~Albarqouni,
  and A.~Mukhopadhyay, ``{GANs} for medical image analysis,'' {\em arXiv
  preprint arXiv:1809.06222}, 2018.

\bibitem{yosinski2014transferable}
J.~Yosinski, J.~Clune, Y.~Bengio, and H.~Lipson, ``How transferable are
  features in deep neural networks?,'' in {\em Advances in Neural Information
  Processing Systems}, pp.~3320--3328, 2014.

\bibitem{simonyan2014very}
K.~Simonyan and A.~Zisserman, ``Very deep convolutional networks for
  large-scale image recognition,'' in {\em International Conference on Learning
  Representations}, 2015.

\bibitem{lin2017focal}
T.-Y. Lin, P.~Goyal, R.~Girshick, K.~He, and P.~Doll{\'a}r, ``Focal loss for
  dense object detection,'' in {\em Proc. IEEE International Conference on
  Computer Vision}, pp.~2980--2988, 2017.

\bibitem{araujo2017classification}
T.~Ara{\'u}jo, G.~Aresta, E.~Castro, J.~Rouco, P.~Aguiar, C.~Eloy,
  A.~Pol{\'o}nia, and A.~Campilho, ``Classification of breast cancer histology
  images using convolutional neural networks,'' {\em PloS one}, vol.~12, no.~6,
  p.~e0177544, 2017.

\bibitem{codella2018skin}
N.~C. Codella, D.~Gutman, M.~E. Celebi, B.~Helba, M.~A. Marchetti, S.~W. Dusza,
  A.~Kalloo, K.~Liopyris, N.~Mishra, H.~Kittler, {\em et~al.}, ``Skin lesion
  analysis toward melanoma detection: A challenge at the 2017 international
  symposium on biomedical imaging ({ISBI}), hosted by the international skin
  imaging collaboration ({ISIC}),'' in {\em Proc. IEEE International Symposium
  on Biomedical Imaging}, pp.~168--172, 2018.

\bibitem{kingma2014adam}
D.~P. Kingma and J.~Ba, ``Adam: A method for stochastic optimization,'' in {\em
  International Conference on Learning Representations}, 2015.

\bibitem{zeiler2012adadelta}
M.~D. Zeiler, ``{ADADELTA}: an adaptive learning rate method,'' {\em arXiv
  preprint arXiv:1212.5701}, 2012.

\bibitem{perez2018data}
F.~Perez, C.~Vasconcelos, S.~Avila, and E.~Valle, ``Data augmentation for skin
  lesion analysis,'' in {\em Proc. International Workshop on Computer-Assisted
  and Robotic Endoscopy}, pp.~303--311, 2018.

\bibitem{esteva2017dermatologist}
A.~Esteva, B.~Kuprel, R.~A. Novoa, J.~Ko, S.~M. Swetter, H.~M. Blau, and
  S.~Thrun, ``Dermatologist-level classification of skin cancer with deep
  neural networks,'' {\em Nature}, vol.~542, no.~7639, p.~115, 2017.

\bibitem{selvaraju2017grad}
R.~R. Selvaraju, M.~Cogswell, A.~Das, R.~Vedantam, D.~Parikh, and D.~Batra,
  ``{Grad-CAM}: Visual explanations from deep networks via gradient-based
  localization,'' in {\em Proc. IEEE International Conference on Computer
  Vision}, pp.~618--626, 2017.

\bibitem{mcinnes2018umap}
L.~McInnes, J.~Healy, and J.~Melville, ``{UMAP}: Uniform manifold approximation
  and projection for dimension reduction,'' {\em The Journal of Open Source
  Software}, 2018.

\end{thebibliography}

\end{document}